\documentclass[twocolumn,superscriptaddress,groupedaddress,amsmath,amssymb,aps,pra,floatfix]{revtex4-2}

\usepackage{comment}
\usepackage{graphicx}
\usepackage{dcolumn}
\usepackage{bm}

\usepackage[utf8]{inputenc}
\usepackage[T1]{fontenc}

\usepackage{braket}
\usepackage{siunitx}
\usepackage{xcolor}

\usepackage[normalem]{ulem}

\def\Rb87{^{87}\mathrm{Rb}}                             
\def\He4{^{4}\mathrm{He}}     

\begin{document}

\title{Spatiotemporal Chaos and Extended Self-Similarity of Bose Einstein Condensates in a 1D Harmonic Trap}

\author{Mingshu Zhao}
\affiliation{Joint Quantum Institute, University of Maryland , College Park, Maryland 20742, USA }
\email{zmshum@terpmail.umd.edu}

\date{\today} 

\begin{abstract}
We investigate spatiotemporal chaos in Bose-Einstein condensate (BEC) confined by a 1D harmonic trap using Gross-Pitaevskii equation simulations. The chaos arises from nonlinear mixing of ground and excited states, confirmed by positive Lyapunov exponents. By sampling the density field at intervals matching the center-of-mass oscillation period, we analyze the density structure function.  Both spatial and temporal density structure functions reveal Kolmogorov-like scaling through extended self-similarity (ESS).  Our findings suggest that ESS and density structure functions provide experimentally accessible tools to explore spatiotemporal chaos and turbulence-like behavior in BECs.
\end{abstract}

\maketitle


\section{Introduction}
\label{sec:intro}

Chaos poses significant challenges due to its inherent unpredictability, even in systems with few degrees of freedom~\cite{lorenz1963deterministic,strogatz2018nonlinear}. Spatiotemporal chaos compounds this complexity, involving systems with infinitely many interacting degrees of freedom~\cite{datseris2022pattern,bvrezinova2011wave}. One classical example is turbulence, which describes the random and chaotic motion of fluids. In turbulence, energy is injected at large scales, transported through the inertial range, and dissipated at small scales~\cite{kolmogorov1941degeneration,kolmogorov1991dissipation,kolmogorov1991local}. A hallmark of the inertial range is the presence of power-law scaling, indicating self-similarity~\cite{kolmogorov1995turbulence}.

However, in many systems—such as turbulence at moderate Reynolds numbers, where a well-defined inertial range is absent—clear scale decomposition is impossible, yet self-similarity persists~\cite{benzi1993extended}. Extended Self-Similarity (ESS) addresses this issue by uncovering scaling laws even without a strict inertial range~\cite{benzi1993extended}. In ESS, the structure function of order $p$, defined as $S_p(r) = \langle |\delta v(r)|^p \rangle$, where $\delta v(r)$ represents the velocity increment over a distance $r$, is compared across different orders. This method reveals correlations that classical approaches often overlook.

This work focuses on atomic Bose-Einstein condensates (BECs), which face greater challenges in turbulence studies compared to classical incompressible fluids due to two key factors: the narrow inertial range and the compressible nature of BEC turbulence\cite{madeira2020quantum}. 
The limited inertial range, spanning less than one order of magnitude in momentum $k$-space, restricts the application of traditional scaling laws~\cite{tanogami2021theoretical}. ESS can potentially address this challenge, as it enables scaling analysis even when a well-defined inertial range is not present.
Moreover, atomic BECs are compressible fluids. A density-weighted velocity field is essential to apply Kolmogorov scaling, which is relevant for vortex turbulence in these systems~\cite{aluie2011compressible, aluie2013scale, tanogami2021theoretical}. However, in systems without dense vortex clusters, wave turbulence—focused on density fluctuations—provides a more accurate description~\cite{zhu2023direct, nazarenko2011wave}. 
In general, BEC turbulence is a hybrid of both vortex and wave turbulence, making it necessary to consider both types of fluctuations.

To investigate scaling behavior and intermittency, it is essential to measure structure functions. In the case of vortex turbulence, velocity structure functions are key and have been experimentally measured recently due to the development of velocity field measurement 
  ~\cite{zhao2024kolmogorov}. For wave turbulence, density structure functions based on density increments offer valuable insights; however, such measurements have only been applied to study instabilities in magneto-optical traps~\cite{griffin2023analysing}, where the atomic clouds are thermal and far from being in a condensate state. Fortunately, density structure functions in BECs can be directly obtained by in-situ absorption imaging. 
For a comprehensive understanding of hybrid turbulence, it is beneficial to measure both density and velocity structure functions.
Currently, most atomic BECs turbulence experiments use time-of-flight imaging, which yields momentum distributions~\cite{thompson2013evidence,navon2016emergence}. Although informative, this approach does not reveal spatial structures in the density field, limiting its utility for studying intermittency and the application of ESS to extend scaling range.

In this work, we investigate the chaotic dynamics of a BEC in a 1D harmonic trap—a simpler model than typical turbulence. On one hand, vortices cannot form in 1D, which makes the system more analogous to wave turbulence. On the other hand we exclude external forcing or dissipation, and the chaotic dynamics simply originates from the nonlinear interaction in the Gross-Pitaevskii equation (GPE)~\cite{pitaevskii2016bose}. Unlike traditional wave turbulence, which typically includes dissipation, this system evolves without dissipation, leading us to describe it as spatiotemporal chaos.

We initialize the system as a superposition of the ground and first excited states, inducing chaotic behavior as nonlinear interactions populate higher excited states with large wave numbers. Through these interactions, populations are redistributed among eigenmodes in a way that resembles an energy cascade from large to small scales. However, without a well-defined inertial range, a central question arises: does the system exhibit self-similarity? Specifically, we aim to determine if the density structure function shows power-law scaling, potentially with ESS.

This phenomenon is expected to be broadly applicable to other trapped BEC systems, but we choose a harmonic trap here for simplicity. According to the Kohn theorem~\cite{kohn1961cyclotron, brey1989optical, dobson1994harmonic}, the center-of-mass (COM) motion in a harmonic trap oscillates at the trap frequency $\omega$, regardless of interaction strength. This behavior allows us to discretize the continuous time evolution by capturing density snapshots at the COM’s oscillation peaks, effectively creating a stroboscopic Poincaré map~\cite{strogatz2018nonlinear}. This approach reduces the system’s continuous evolution to a discrete sequence, preserving key dynamical features and enabling us to decompose the density field into mean and fluctuating components for structure function analysis.

With this setup, we aim to explore scaling behavior and identify potential self-similarity within the system’s chaotic dynamics. Specifically, we analyze correlations in the spatiotemporal chaos using two types of density increments: spatial density increments, $\delta \rho(r)$, over a spatial separation $r$; and temporal density increments, $\delta \rho(t)$, over a time interval $t$.

Although our system does not exhibit classical turbulence due to the absence of an inertial range, we employ ESS to reveal underlying scaling behavior. By comparing $p$th-order structure functions to $S_3$ (the third-order structure function), we observe scaling exponents that approximate $p/3$ for both spatial and temporal density structure functions. This scaling is consistent with the K41 scalings based on self-similarity hypothesis~\cite{kolmogorov1941degeneration,kolmogorov1991local,kolmogorov1991dissipation}, even though the typical conditions for turbulence are not strictly present. The alignment with K41-like exponents in our ESS analysis suggests that, despite lacking an inertial range, the system’s density fluctuations follow a self-similar pattern.
 These findings highlight that the density structure function with ESS serves as a powerful tool for investigating spatiotemporal chaos and turbulence in BECs.



The remainder of this paper is organized as follows: In Section~\ref{sec:methods}, we introduce the system and outline the methods used in our numerical experiments. Section~\ref{sec:chaos} focuses on analyzing the chaotic behavior of the system, including sensitivity to initial conditions and the presence of positive Lyapunov exponents~\cite{strogatz2018nonlinear}. In Section~\ref{sec:spatial}, we examine spatial correlations through spatial density structure functions, applying both direct analysis and ESS, where K41 scaling emerges under ESS. We also propose an experiment setup to validate our findings. Section~\ref{sec:temporal} investigates temporal correlations using temporal density structure functions, with K41 scaling again observed under ESS. Finally, in Section~\ref{sec:conclude}, we summarize our findings and discuss potential future research directions.


\section{Methods}
\label{sec:methods}
\subsection{Gross-Pitaevskii Equation}
The dynamics of a zero temperature weakly interacting BEC in a 1D harmonic trap are described by the time-dependent Gross-Pitaevskii Equation:
\begin{equation}
i\hbar \frac{\partial \tilde{\Psi}(\tilde{x}, \tilde{t})}{\partial \tilde{t}} = 
\left( -\frac{\hbar^2}{2m} \frac{\partial^2}{\partial \tilde{x}^2} 
+ \frac{1}{2} m \omega^2 \tilde{x}^2 + {g}_{\text{1D}} |\tilde{\Psi}(\tilde{x}, \tilde{t})|^2 \right) \tilde{\Psi}(\tilde{x}, \tilde{t}),
\end{equation}
where $\tilde{\Psi}(\tilde{x}, \tilde{t})$ is the macroscopic wavefunction, $\tilde{x}$ and $\tilde{t}$ are the spatial and temporal coordinates, $m$ is the particle mass, $\omega$ is the trap frequency, and ${g}_{\text{1D}}$ is the interaction strength. The harmonic potential $V(\tilde{x}) = \frac{1}{2} m \omega^2 \tilde{x}^2$ confines the condensate along the 1D axis.

The wavefunction must satisfy the following normalization condition:
\begin{equation}
\int_{-\infty}^{\infty} |\tilde{\Psi}(\tilde{x}, \tilde{t})|^2 \, d\tilde{x} = N,
\end{equation}
where $N$ is the total number of particles in the condensate.

To work in dimensionless units, we introduce the following transformations:
\begin{equation}
x = \tilde{x} \sqrt{\frac{m \omega}{\hbar}}, \quad 
t = \omega \tilde{t}, \quad 
\psi(x, t) = \tilde{\Psi}(\tilde{x}, \tilde{t}) \left(\frac{m \omega}{\hbar}\right)^{-1/4}N^{-1/2},
\end{equation}
where $x$, $t$, and $\psi(x, t)$ are the dimensionless quantities used in the rest of the paper. Under this transformation, the relation between the dimensionless interaction strength $g$ and the dimensional interaction strength $\tilde{g}_{\text{1D}}$ is:
\begin{equation}
g = \frac{{g}_{\text{1D}} Nm^{1/2}}{\hbar^{3/2}\omega^{1/2}}.
\end{equation}

The GPE in dimensionless form becomes:
\begin{equation}
i \frac{\partial \psi(x, t)}{\partial t} = 
\left( -\frac{1}{2} \frac{\partial^2}{\partial x^2} + \frac{1}{2} x^2 + g |\psi(x, t)|^2 \right) \psi(x, t),
\end{equation}
which satisfies the following normalization condition:
\begin{equation}
\int_{-\infty}^{\infty} |\psi(x, t)|^2 \, dx = 1,
\end{equation}

The stationary solution of the GPE takes the form
\begin{equation}
    \psi(x, t) = e^{-i\mu (t - t_0)} \psi(x, t_0),
\end{equation}
where $\mu$ is the chemical potential of the system, and $t_0$ is the initial time. Substituting this ansatz into the GPE transforms it into the following time-independent form:
\begin{equation}
    \left( -\frac{1}{2} \frac{d^2}{dx^2} + \frac{1}{2} x^2 + g |\psi(x)|^2 \right) \psi(x) = \mu \psi(x).
\end{equation}
Here, $\mu$ plays the role of the eigenenergy in this nonlinear Schrödinger equation, determining the chemical potential associated with each stationary state.

We denote the ground state and excited states of the system as $\psi_0(x)$, $\psi_1(x)$, $\psi_2(x), \dots$, with corresponding eigenenergies $\mu_0$, $\mu_1$, $\mu_2, \dots$. 
For simplicity we treat the stationary wavefunction as real-valued function.
And we will use the superposition of the ground and first excited state as the initial state in this work.
The ground state $\psi_0(x)$ corresponds to the lowest eigenenergy $\mu_0$ and has even parity, while the first excited state $\psi_1(x)$ has odd parity and corresponds to the eigenenergy $\mu_1$.

\subsection{Numerical Methods}

To study the dynamics of the GPE, we employ the time-splitting Crank-Nicholson finite difference method for both imaginary and real-time evolution. Imaginary time evolution is used to compute the ground and first excited states: the ground state, $\psi_0(x)$, is obtained by evolving an even initial wavefunction, while the first excited state, $\psi_1(x)$, is obtained by evolving an odd initial wavefunction. The spatial grid consists of 4096 points, spanning the dimensionless range $x \in [-16, 16]$. Details of the numerical scheme are provided in Appendix~\ref{app:CNFD}.

\begin{figure*}[htb!]
\centering
\includegraphics{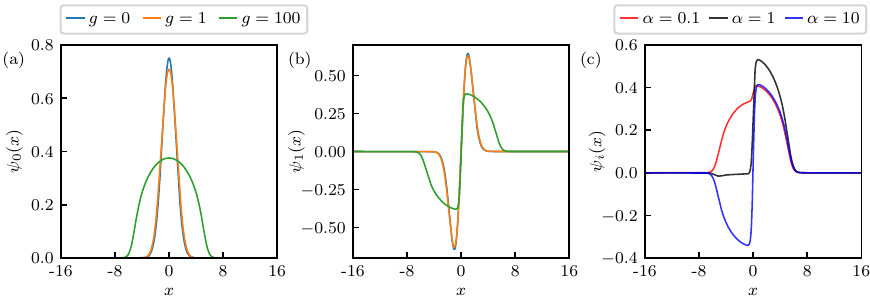}
\caption{(a) Ground state $\psi_0(x)$. (b) First excited state $\psi_1(x)$. (c) Iniital wavefunction $\psi_i(x)$ for real time evolution when $g=100$.}
\label{fig:initial_states}
\end{figure*}

Figure~\ref{fig:initial_states} presents the ground state $\psi_0(x)$ in panel (a) and the first excited state $\psi_1(x)$ in panel (b) for interaction strengths $g = 0, 1,$ and $100$. The case with $g = 100$, and a trap frequency $\omega=2\pi\times20\,\mathrm{Hz}$ for $^{87}\mathrm{Rb}$, results in a speed of sound $c \approx 1.6\,\mathrm{mm/s}$, which is comparable to typical experimental values~\cite{zhao2024kolmogorov, andrews1997propagation}.

The initial wavefunction for real-time evolution is constructed as: 
\begin{equation} 
\psi_i(x) = \frac{\psi_0(x) + \alpha \psi_1(x)}{\sqrt{1+\alpha^2}}, 
\end{equation} 
where $\alpha$ controls the contribution of the first excited state. Figure~\ref{fig:initial_states}(c) shows the initial states for three different values of $\alpha$:
(i) For $\alpha = 0.1$ (red), there is a small mixing of $\psi_1$ with $\psi_0$, resulting in no phase jump.
(ii) For $\alpha = 1$ (black), there is a balanced mixing of $\psi_0$ and $\psi_1$, which introduces a $\pi$ phase jump and gives an asymmetric density profile. (iii) For $\alpha = 10$ (blue), there is a small mixing of $\psi_0$ into $\psi_1$, with the $\pi$ phase jump located close to the origin, resulting in a profile that closely resembles a dark soliton~\cite{burger1999dark}.


\begin{table}[htb]
\caption{\label{tab:eigenenergy}%
Values of $\mu_0$, $\mu_1$, and $\Delta \mu = \mu_1 - \mu_0$ for different interaction strengths $g$.}
\vspace{0.5cm}
\begin{ruledtabular}
\begin{tabular}{c d d d}
\textrm{$g$} & 
\multicolumn{1}{c}{$\mu_0$} & 
\multicolumn{1}{c}{$\mu_1$} & 
\multicolumn{1}{c}{$\Delta \mu$} \\
\hline
0 & 0.5000 & 1.5000 & 1.0000 \\ 
0.1 & 0.5396 & 1.5298 & 0.9902 \\
1 & 0.8699 & 1.7901 & 0.9202 \\
10 & 3.1072 & 3.8632 & 0.7560 \\ 
100 & 14.1343 & 14.8504 & 0.7161 \\
\end{tabular}
\end{ruledtabular}
\end{table}

Table~\ref{tab:eigenenergy} summarizes the eigenenergies $\mu_0$ and $\mu_1$, along with their difference $\Delta \mu = \mu_1 - \mu_0$, for various values of the dimensionless interaction strength $g$. Notably, $\Delta \mu$ becomes smaller than the Kohn mode frequency $\omega = 1$ when interactions are introduced. This value of $\Delta \mu$ aligns closely with the oscillation frequency of a dark soliton. Under the Thomas-Fermi approximation, this frequency is approximately $\sqrt{2}$ times the Kohn mode frequency~\cite{becker2008oscillations}, which matches our results for $g = 100$.

In the subsequent analysis, we primarily focus on the cases $\alpha = 0.1$ and $\alpha = 1$ with $g = 100$.

\section{Chaos Analysis}
\label{sec:chaos}
\subsection{Center-of-mass Motion and Nonlinear Mixing}
In the harmonic trap, the Kohn theorem guarantees that the COM motion of the system oscillates with the trap frequency $\omega$, independent of the interaction strength.
A short proof can be found in the Appendix~\ref{app:Kohn}.

Now we consider the evolution of $\psi_i$. 
For noninteracting case ($g=0$), the state evolves to 
\begin{equation}
    \psi(x,t)= \frac{\psi_0(x)e^{-i\mu_0 t}+\alpha\psi_1(x)e^{-i\mu_1 t}}{\sqrt{1+\alpha^2}},
\end{equation}
which gives the density field
\begin{equation}
    \rho(x,t) =\frac{\psi^2_0(x)+\alpha^2\psi^2_1(x)+2\alpha\psi_0(x)\psi_1(x)\cos[(\mu_1-\mu_0)t]}{1+\alpha^2}.
\end{equation}
The whole system undergoes periodic motion with an angular frequency 
$\mu_1 - \mu_0 = 1$, as shown in Table~\ref{tab:eigenenergy}, which is consistent 
with the Kohn theorem.

However, when nonlinear interactions are introduced, Eqs.~(10)–(11) no longer apply, otherwise the system would oscillate at a frequency smaller 
than what the Kohn theorem predicts (see Table~\ref{tab:eigenenergy}). This highlights the 
crucial role of nonlinear interactions in the dynamics, as they redistribute energy 
among modes, generating higher excited states with large $k$-modes. This behavior 
resembles the kinetic energy cascade in 3D turbulent incompressible fluids, 
where energy flows from large to small scales within the inertial range.
Nonetheless, this analogy with turbulence is purely phenomenological, as our system lacks the well-defined inertial range necessary for classical turbulence. 
Despite this, we will show next the system undergoes chaotic dynamics.

\subsection{Poincaré Section and Discrete Density Field Sequence}

To simplify the analysis, we discretize the wavefunction sequence by recording the wavefunction each time the COM reaches a local maximum. 
This approach is analogous to a Poincaré map, a technique in chaos studies that reduces continuous dynamics to discrete snapshots. By capturing the system’s state each time it crosses a particular phase section, the Poincaré map provides a compact representation of the trajectory in phase space. This technique is widely used to reveal both periodic and aperiodic (chaotic) behavior, including the complex, fractal structures seen in strange attractors.

\begin{figure*}[htb!]
\centering
\includegraphics{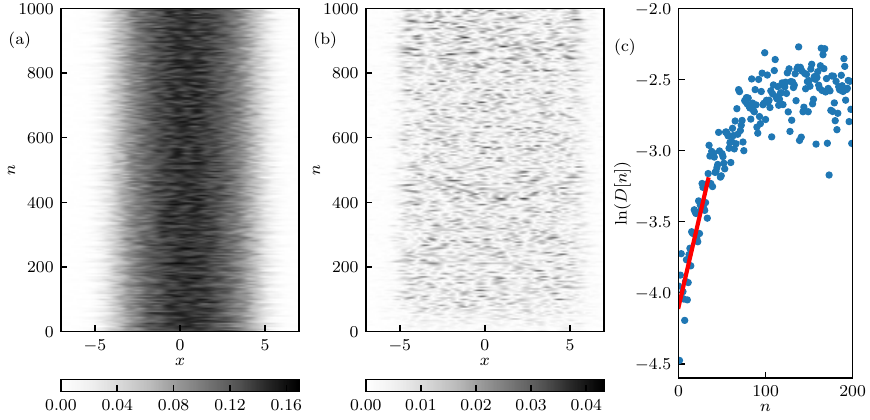}
\caption{(a) Discrete density sequence $\rho_n(x)$ for $g=100$ and $\alpha=0.1$. (b) Sequence of density differences $|\rho_{n, \alpha=0.1} - \rho_{n, \alpha=0.09}|$ for $g=100$, illustrating divergence due to initial perturbation. (c) Plot of the logarithmic distance sequence $\ln(D[n])$ from panel (b) as a function of $n$ (blue dots), where $D[n] = D[\rho_{n, \alpha=0.1}(x), \rho_{n, \alpha=0.09}(x)]$. The red line, fitted over the first $35$ cycles, gives a Lyapunov exponent of $\lambda = 0.026(2)$.
}
\label{fig:Lyapunov}
\end{figure*}

The Kohn theorem ensures that COM oscillates at the trap frequency, $\omega = 1$. Consequently, we capture wavefunction snapshots at intervals of $2\pi$, corresponding to each complete oscillation cycle. Figure~\ref{fig:Lyapunov}(a) shows the resulting density profiles for the discrete sequence 
$$\rho_n(x) = \rho(x, t = 2\pi n), \quad n = 0, 1, 2, \dots,$$ 
simulated over 1000 cycles for $g = 100$ and $\alpha = 0.1$. Initially, the density profile exhibits smooth patterns dominated by low spatial frequency $k$ fluctuations. However, as time progresses, the profile transitions into a more complex, aperiodic structure with increased high-$k$ fluctuations, providing a qualitative indication of spatiotemporal chaos.

\subsection{Lyapunov Exponent}

The chaotic behavior of the system can be quantitatively analyzed by perturbing the initial condition and observing how two initially close trajectories diverge over time. In this setup, we introduce a slight perturbation to the initial parameter $\alpha$ of the state $\psi_i$, allowing us to track the effect of this small change on the evolution of the density profiles and the separation of the resulting trajectories.

To quantify this divergence, we define the distance between two density profiles using the $L^1$-norm:
\begin{equation}
    D[\rho_A(x), \rho_B(x)] = \int_{-\infty}^{+\infty} dx \, 
    \left| \rho_A(x) - \rho_B(x) \right|,
\label{eq:L1_norm}
\end{equation}
where $\rho_A(x)$ and $\rho_B(x)$ denote the density profiles of two states evolving from slightly different initial conditions.

The Lyapunov exponent $\lambda$ provides a measure of the average rate at which two close trajectories diverge in phase space. In our analysis, we consider two sequences of density profiles, $[\rho_{n, \alpha}(x)]$ and $[\rho_{n, \alpha + \Delta \alpha}(x)]$, where the initial state parameter $\alpha$ is perturbed by a small amount $\Delta \alpha$. For each step $n$, we define the distance sequence as:
$$
D[n] = D[\rho_{n, \alpha}(x), \rho_{n, \alpha + \Delta \alpha}(x)].
$$
The Lyapunov exponent $\lambda$ is then estimated by fitting the growth of $D[n]$ to an exponential form:
\begin{equation}
    D[n] \propto d \, e^{\lambda n},
\end{equation}
where $d$ is the initial separation. A positive Lyapunov exponent, $\lambda > 0$, indicates chaotic behavior, as trajectories diverge exponentially over time. In contrast, $\lambda = 0$ or negative values imply stable or non-chaotic evolution.

An example of this divergence is illustrated in Fig.~\ref{fig:Lyapunov}(b), showing the density difference between $\alpha = 0.1$ and $\alpha = 0.09$. Over time, a marked increase in this density difference is observed. Panel (c) of the figure shows the sequence of distances $D[n]$, with the red line representing a fitted slope that gives the Lyapunov exponent.


\begin{table*}[htb!]
\vspace{0.5cm}
\caption{\label{tab:Lyapunov}%
Lyapunov exponents $\lambda$ for different $g$ and $\alpha$.}
\vspace{0.5cm}
\begin{ruledtabular}
\begin{tabular}{c c c c c c c c}
\textrm{} & 
\multicolumn{3}{c}{$g = 0.1$} & 
\multicolumn{3}{c}{$g = 1$} \\
\cline{2-7}
$\alpha$ & 
\textrm{0.1} & 
\textrm{1} & 
\textrm{10} & 
\textrm{0.1} & 
\textrm{1} & 
\textrm{10} \\
\hline
$\lambda$ & $-7(9) \times 10^{-7}$ & $9(2) \times 10^{-5}$ & $5(6) \times 10^{-5}$ & 
$1(3) \times 10^{-6}$ & $2.2(2) \times 10^{-2}$ & $2.9(6) \times 10^{-4}$ \\
\hline
& \multicolumn{3}{c}{$g = 10$} & \multicolumn{3}{c}{$g = 100$} \\
\cline{2-7}
$\alpha$ & 
\textrm{0.1} & 
\textrm{1} & 
\textrm{10} & 
\textrm{0.1} & 
\textrm{1} & 
\textrm{10} \\
\hline
$\lambda$ & $4.8(9) \times 10^{-3}$ & $1.1(1) \times 10^{-1}$ & $4.8(6) \times 10^{-3}$ & 
$2.6(2) \times 10^{-2}$ & $2.2(4) \times 10^{-1}$ & $3.3(1) \times 10^{-2}$ \\
\end{tabular}
\end{ruledtabular}
\end{table*}

Table~\ref{tab:Lyapunov} presents the calculated Lyapunov exponents $\lambda$ for various $g$ and $\alpha$ values. We observe that $\lambda$ increases with higher interaction strengths $g$, and the system becomes chaotic for $g = 10$ and $g = 100$, regardless of $\alpha$. Furthermore, the balanced mixture of the ground and first excited states ($\alpha = 1$) consistently shows stronger chaotic behavior compared to more imbalanced mixtures ($\alpha = 0.1, 10$).

\section{Spatial Density Structure Function Analysis}
\label{sec:spatial}

The positive Lyapunov exponent shown in the previous section confirms that our system exhibits spatiotemporal chaos. In this section, we focus on investigating the spatial structure of this chaotic behavior. Spatiotemporal chaos in systems like turbulent fluid flow is often analyzed through structure functions and probability density functions (PDFs), which help reveal scaling laws and self-similar behavior. We apply similar techniques here to determine whether our system displays analogous scaling behavior and self-similarity.

\subsection{Spatial Density Structure Functions}

\begin{figure*}[htb!]
\centering
\includegraphics{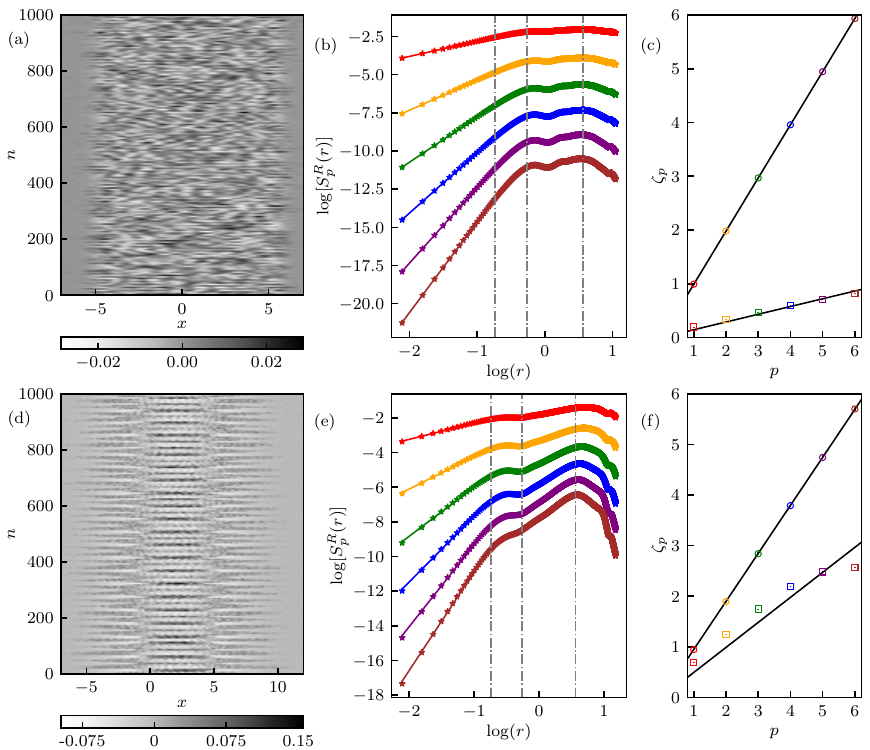}
\caption{
Top panel: results for $g=100$, $\alpha=0.1$; bottom panel: results for $g=100$, $\alpha=1$. 
(a, d) Fluctuating density sequence $\rho_n'(x)$. 
(b, e) Log-log plot of spatial density structure function $S_p^R(r)$ versus $r$ for $p=1 \dots 6$, with color coding as shown on the x-axis in panels (c, f). 
Grey dashed lines indicate positions of $r_0$, $r_1 = 3r_0$, and $r_2 = 20r_0$, respectively. 
(c, f) Scaling exponent $\zeta_p$ versus $p$, where $\zeta_{p1}$ (circle markers) is extracted from region 1 ($r < r_0$) and $\zeta_{p2}$ (square markers) is extracted from region 2 ($r_1 < r < r_2$). 
The black line represents a linear fit.
}
\label{fig:struc_func}
\end{figure*}

We begin by decomposing the discrete density sequence into its mean and fluctuating components, similar to the Reynolds decomposition~\cite{reynolds1895iv} in turbulence studies, where the flow is separated into mean and fluctuating parts. For the density field, this decomposition is expressed as:
\begin{equation}
    \rho_n(x) = \langle \rho(x) \rangle + \rho'_n(x),
\end{equation}
where $\langle \rho(x) \rangle$ represents the mean density field, averaged over all snapshots in the sequence, and $\rho'_n(x)$ denotes the fluctuations around the mean at each time step $n$. As an example, Fig.~\ref{fig:struc_func}(a) and (d) display $\rho'_n(x)$ for the cases of $g=100$ with $\alpha=0.1$ and $\alpha=1$, respectively.

The spatial structure of the density fluctuations is analyzed through the spatial density increment, defined as $\delta \rho(r) = \rho'_n(x + r) - \rho'_n(x)$. From these increments, we can study statistical properties such as the PDF and the structure function defined below.

For a given spatial displacement $r$, the PDF of spatial density increments is obtained from a histogram, represented as $P[\delta \rho(r)]$. Next, we calculate the spatial density structure function of order $p$, which is defined as:
\begin{equation}
    S_p^R(r) = \int_{-\infty}^{+\infty} \left| \delta \rho(r) \right|^p \, P[\delta \rho(r)] \, d[\delta \rho(r)],
\label{Eq:struc_func}
\end{equation}
where the superscript $R$ indicates the spatial structure function.

Structure functions are widely used in turbulence studies because they reveal how fluctuations vary with scale. For systems with self-similar behavior, the structure function follows a power-law scaling:
\begin{equation}
    S_p(r) \propto r^{\zeta_p},
\end{equation}
where $\zeta_p$ is the scaling exponent. Figures~\ref{fig:struc_func}(b) and (e) display the spatial density structure functions from $p=1$ to $p=6$, calculated from the fluctuating density fields shown in panels (a) and (d). The vertical dashed lines mark $r_0$, $r_1 = 3r_0$, and $r_2 = 20r_0$, where $r_0 = 1/\sqrt{2\mu_0}$ represents the ground-state healing length.

We observe a clear power-law scaling within the region $r < r_0$, and the scaling $\zeta_{p1}$, extracted from the linear fit in the log-log plot in panels (b) and (e), is shown with circular markers in panels (c) and (f). The scaling exponent $\zeta_{p1}/p$ remains close to 1 across different values of $g$ and $\alpha$.
This scaling behavior can be understood by considering the Taylor expansion: for small spatial displacement $r$, if the density fluctuation field is differentiable, then $|\delta \rho(r)|^p \approx |(\partial \rho / \partial x)|^p r^p$, which explains why $\zeta_p/p$ is approximately 1 for all cases.

In the range $r_1 < r < r_2$, we observe another power-law behavior. The scaling exponent in this region, $\zeta_{p2}$, is shown with square markers in panels (c) and (f). For the weak initial mixing case ($\alpha = 0.1$), $\zeta_{p2}$ remains linear with $p$, while for the strong mixing case ($\alpha = 1$), $\zeta_{p2}$ is larger and deviates from linearity, suggesting intermittency. 

Although power-law scaling is observed in these two regions, the range $r < r_0$ is typically difficult to measure experimentally and is of limited physical interest. The second region, $r_1 < r < r_2$, is more accessible experimentally, but the scaling range of $r_2/r_1 \approx 6.7$ is not large enough for conclusive results. Thus, applying ESS becomes necessary for more reliable power-law extraction.



\subsection{Extended Self-Similarity}

ESS is a technique used in turbulence and complex systems 
to extract scaling laws more effectively. In traditional turbulence studies, the structure 
function of order $p$, $S_p(r)$, is analyzed as a function of the spatial separation $r$. 
However, finding a well-defined scaling range in real-world data can be challenging, 
especially when the inertial range is poorly developed.

ESS addresses this by comparing structure functions of different orders against 
each other, rather than directly using the separation distance $r$. Specifically, in our case, ESS 
plots $S_p^R(r)$ as a function of a reference structure function, often $S_3^R(r)$, which 
corresponds to the third-order structure function:
\begin{equation}
    S_p^R(r) \propto S_3^R(r)^{\beta_p}.
\end{equation}
Here, $\beta_p$ are scaling exponents determined through these inter-order comparisons. 
This approach reveals scaling behavior even when the classical power-law 
dependence on $r$ is obscured.

\begin{figure*}[htb!]
\centering
\includegraphics{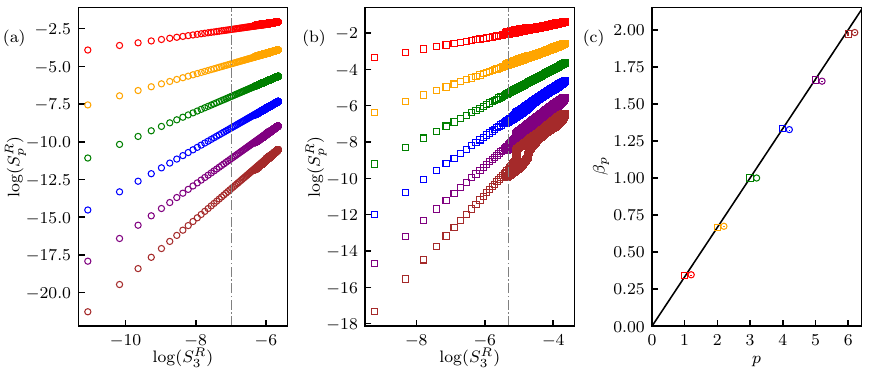}
\caption{(a) Log-log plot of $S_p^R$ versus $S_3^R$ for $g=100$, $\alpha=0.1$. The color coding for each $p$ value is shown on the horizontal axis in panel (c). The grey dashed line indicates $S_3^R(r_0)$. 
(b) Log-log plot of $S_p^R$ versus $S_3^R$ for $g=100$, $\alpha=1$, with the grey dashed line marking $S_3^R(r_0)$. 
(c) ESS scaling exponents $\beta_p$, obtained from linear fits to the data in (a) (circle markers) and (b) (square markers). For clarity, the circle markers are shifted to the right by $0.2$ to avoid overlap. The black line represents the K41 scaling law, $p/3$. }
\label{fig:ESS}
\end{figure*}
For example, in Fig.~\ref{fig:ESS}(a), we plot the complete dataset from Fig.~\ref{fig:struc_func}(b) for the weak mixing case ($\alpha=0.1$), where the grey dashed line corresponds to $S_3^R(r_0)$. We observe that ESS extends across the full spatial range, even when the structure functions exhibit different scalings with respect to $r$ across spatial scales.
Similarly, Fig.~\ref{fig:ESS}(b) shows the dataset from Fig.~\ref{fig:struc_func}(e) for the strong mixing case ($\alpha=1$). Again, ESS spans the entire spatial range, although regions $r > r_2$ and $r_0 < r < r_2$ display slight deviations, as seen in the $p=6$ curve (brown), which bifurcates to the right of the grey dashed line. 
ESS is further reflected in the scaling exponent $\beta_p$.
Since ESS uses the third-order structure function as a reference, the expected scaling $\beta_p$ for both strong and weak mixing approximates the K41 scaling $p/3$, as shown in Fig.~\ref{fig:ESS}(c).
For those interested in the intermittency, please refer to Appendix~\ref{app:intermittency}, where we fit deviations using KO62 theory~\cite{kolmogorov1962refinement} to obtain the intermittency exponent.

In summary, we confirm that ESS is present and provides a valuable means to extend the experimental scaling range for studying power laws.

\subsection{Probablity Density Function}

Despite substantial research on ESS in turbulence, its underlying mechanisms remain unclear and can only be derived under certain assumptions~\cite{yakhot2001mean,ching2002extended,chakraborty2010extended}. 
Here, we do not aim to mathematically prove the existence of ESS in our case. Rather, our objective is to explore how ESS can be interpreted through the PDF, as all orders of the structure function, according to Eq.~\eqref{Eq:struc_func}, depend solely on the PDF $P[\delta\rho(r)]$.
In other words, if $P[\delta\rho(r)]$ exhibits self-similarity across different $r$ values, we expect ESS to hold.
\begin{figure}[htb!]
\centering
\includegraphics{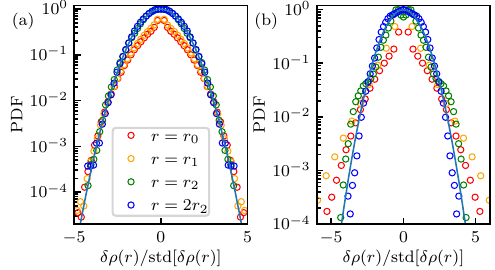}
\caption{Histogram of $\delta\rho(r)$. The maximum bin count is normalized to 1, and $\delta\rho(r)$ is normalized by its standard deviation. The light blue curve represents a Gaussian distribution for comparison. (a) Results for $g=100$, $\alpha=0.1$. (b) Results for $g=100$, $\alpha=1$.}
\label{fig:PDF}
\end{figure}

Figure~\ref{fig:PDF} shows the PDF for both weak and strong mixing cases. In the weak mixing case, shown in panel (a), we observe that although the central region ($|\delta\rho(r)/\mathrm{std}[\delta\rho(r)]|<3$) deviates from Gaussian for small $r$ (red and yellow curves), the distribution in the tail region ($|\delta\rho(r)/\mathrm{std}[\delta\rho(r)]|>3$) is closer to Gaussian, where $\mathrm{std}[\Delta\rho(r)]$ stands for the standard deviation of the density increments. Since the tail behavior primarily influences higher-order structure functions, we infer that the system exhibits a hierarchical structure~\cite{ching2002extended}, which could account for the presence of ESS.

In the strong mixing case, shown in panel (b), the PDF is distinctly non-Gaussian at smaller separations (red, yellow, and green curves). At larger separations (blue curve), the PDF approaches a Gaussian form, resembling intermittency observed in turbulence. As $r$ decreases, the PDF deviates further from Gaussian, exhibiting fatter tails. 
This behavior implies a strong deviation from K41 scaling based on classical turbulence interpretations, as the pronounced non-Gaussian features in the PDF are typically associated with greater intermittency. However, we have observed that this only leads to negligible intermittency in our case, as shown in Fig.~\ref{fig:ESS}(c). This discrepancy suggests a need for deeper insight into the relationship between PDF shape and scaling behavior in this system.

While we cannot fully explain the ESS in terms of the PDFs for the strong mixing case, it is possible that these fat-tailed distributions share a form of self-similarity, with the Gaussian distribution emerging as a specific instance within this family. 
Thus, our understanding of ESS remains qualitative, and a more complete explanation is left for future investigation.

\subsection{Experiment Proposal}

Typical density measurement techniques for BECs, such as absorption imaging, are inherently destructive. Even non-destructive methods—such as partial transfer absorption imaging~\cite{ramanathan2012partial}, phase-contrast imaging~\cite{bradley1997bose,andrews1997propagation}, and Faraday imaging~\cite{gajdacz2013non}—can still perturb the dynamics during a single experimental run, which is not ideal when investigating chaotic behavior.
Therefore, building an ensemble through repeated independent experiments is preferable for statistical analysis. 

\begin{figure*}[hbt!]
\centering
\includegraphics{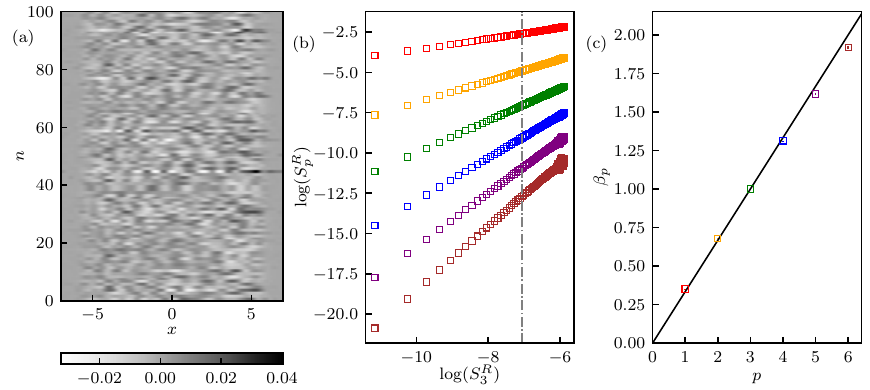}
\caption{(a) Fluctuating density sequence $\rho_n'(x)$ for $g=100$, with $\alpha$ sampled from a normal distribution $\mathcal{N}(0.1, 0.02)$. Here, $n$ represents different realizations of the numerical simulations. (b) Log-log plot of $S_p^R$ versus $S_3^R$ for the data in panel (a), with the grey dashed line indicating $S_3^R(r_0)$. (c) ESS scaling exponents $\beta_p$, determined from linear fits to the data in panel (b). The black line shows the K41 scaling law, $p/3$.}

\label{fig:exp}
\end{figure*}

The initial state can be prepared using a digital micromirror device (DMD), which allows for precise pattern generation~\cite{gauthier2016direct,zhao2024kolmogorov}. 
For simplicity, we consider the case $\alpha = 0.1$ where no $\pi$ phase jump appears in the initial state. The procedure involves evaporating the atomic cloud to form a BEC while projecting a patterned potential over a harmonic trap, ensuring the initial state has zero phase gradient. After preparing this state, the projected potential is abruptly turned off, leaving only the harmonic trap, effectively creating a sudden quench from the DMD-imposed pattern to a pure harmonic potential. Here, we assume that any noise in the system originates from the initial state preparation.

In our numerical simulations, we set $\alpha \sim \mathcal{N}(0.1, 0.02)$, following a normal distribution with a mean of $0.1$ and a standard deviation of $0.02$, across $100$ trials to replicate the uncertainty in preparation.
With the speed of sound in typical experiments around $1 \, \text{mm/s}$, the corresponding interaction strength $g$ is approximately $100$, which we also use in our simulations.

Figure~\ref{fig:exp}(a) displays the ensemble of fluctuating density profiles $\rho_n'(x)$. Panel (b) shows the structure function obtained under ESS, and panel (c) presents the ESS scaling. We find that the ESS scaling aligns with the K41 scaling of $p/3$, indicating that ensemble averaging of the structure function produces the same ESS scaling as time averaging.
Therefore, the ESS with power law scaling is experimentaly observable.

\section{Temporal Density Structure Functions}
\label{sec:temporal}
In this section, we examine the temporal structure of the spatiotemporal chaos. We begin by investigating whether the system tends toward thermalization, transitioning from its initial non-equilibrium state to an equilibrium state. We then extend the structure function analysis to the temporal domain to explore temporal correlations. Remarkably, we observe extended self-similarity in the temporal density structure function.


\subsection{Distance to Average Density}

We first calibrate the thermalization by studying the distance $\bar{D}_n$ between the density $\rho_n(x)$ and the average density $\bar{\rho}(x)$, i.e.,
\begin{equation}
  \bar{D}_n = \int_{-\infty}^{\infty} dx\,\left| \rho_n(x) -\bar{\rho}(x)\right|.
\end{equation}
If the system thermalizes, then $\bar{\rho}(x)$ represents the equilibrium density profile, so we expect $\bar{D}_n$ to decrease from its initial, non-equilibrium state (characterized by a large distance from equilibrium) and eventually fluctuate around a stable value with relatively small amplitude. 

\begin{figure}[htb!]
\centering
\includegraphics{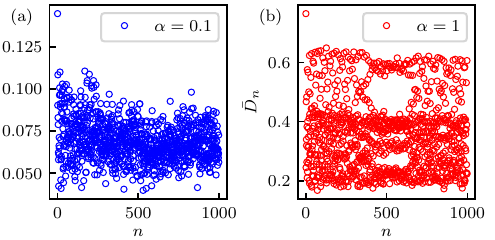}
\caption{Density distance sequence $\bar{D}_n$. (a) Case with $g=100$, $\alpha=0.1$. (b) Case with $g=100$, $\alpha=1$.}
\label{fig:excited_thermalize}
\end{figure}
This behavior is observed in Fig.~\ref{fig:excited_thermalize}(a), where the system thermalizes after approximately 300 cycles under a small initial mixing of $\alpha = 0.1$ for $g = 100$. 
However, the dynamics change significantly for the case of initial balanced mixing, $\alpha = 1$, as shown in panel (b). 
Here, the system starts far from equilibrium and does not appear to thermalize over 1000 cycles, suggesting either a prethermal state or persistent non-equilibrium dynamics with strong intermittent fluctuations analogous to fully developed turbulence.

\subsection{Temporal Density Structure Function}


The temporal density increment with time displacement $n$ (in cycles) is defined using the $L^1$-norm from Eq.~\eqref{eq:L1_norm}:
\begin{equation}
    \delta \rho(n) = \langle D[\rho_i(x), \rho_{i+n}(x)] \rangle,
\end{equation}
where $\langle \cdots \rangle$ denotes averaging over all possible values of $i$. 

Using these temporal density increments, we define the temporal structure function:
\begin{equation}
    S_p^t(n) = \langle |\delta \rho(n)|^p \rangle.
\end{equation}

\begin{figure*}[hbt!]
\centering
\includegraphics{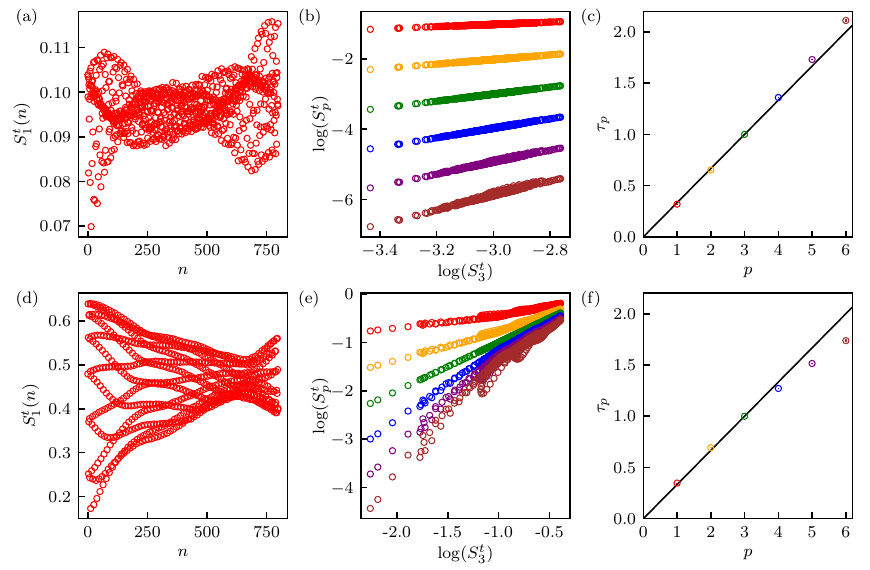}
\caption{Temporal structure functions $S_p^t$ and scaling exponent $\tau_p$ under ESS. The top panel shows the weak mixing case ($g=100$, $\alpha=0.1$), and the bottom panel shows the strong mixing case ($g=100$, $\alpha=1$). (a,d) First-order temporal structure function $S_1^t(n)$. (b,e) Log-log plot of $S_p^t$ versus $S_3^t$ with color coding as indicated on the horizontal axis of the right panel. (c,f) Scaling exponent $\tau_p$, with the black line representing the K41 scaling law.}
\label{fig:Sp_thermalize}
\end{figure*}
Figure~\ref{fig:Sp_thermalize}(a) displays $S_1^t(n)$ for $g=100$ and $\alpha=0.1$. The structure function saturates after approximately 300 cycles, indicating the time scale over which the system loses memory of its initial conditions. Panel (b) shows the higher-order temporal structure functions compared to $S_3^t$ under ESS. A clear power-law relationship is observed, and the scaling $\tau_p$, determined from $S_p^t(n) \propto S_3^t(n)^{\tau_p}$, in panel (c) aligns well with K41 scaling, further confirming self-similar behavior in the temporal dynamics of the system.

In contrast, Fig.~\ref{fig:Sp_thermalize}(d) displays the structure function for $g = 100$ with a higher initial mixing of $\alpha = 1$. The $S_1^t(n)$ reveals a complex bifurcation pattern, which reflects the aperiodic stripe patterns seen in Fig.~\ref{fig:struc_func}(d). Panels (e) and (f) show clear signs of intermittency, presenting a stark contrast to the behavior observed with weak initial mixing ($\alpha = 0.1$) in panel (b-c). This sensitivity to initial conditions highlights the influence of the starting configuration on the dynamics and thermalization process, consistent with the observations in Fig.~\ref{fig:excited_thermalize}.

Interestingly, despite the multi-valued nature of the structure function (due to bifurcations), the power-law scaling in the ESS persists. Intermittency are again observed in the case of strong initial mixing. Details on the intermittency behavior are provided in Appendix~\ref{app:intermittency2} for interested readers.

\section{Conclusion}
\label{sec:conclude}



In this work, we investigated spatiotemporal chaos in a Bose-Einstein condensate confined in a 1D harmonic trap using Gross-Pitaevskii equation simulations. The chaotic behavior arises from nonlinear mixing between the ground state and excited states. We demonstrated the presence of chaos by confirming positive Lyapunov exponents.

To explore the spatial statistics of this chaotic system, we analyzed density structure functions with spatial displacements, observing Kolmogorov-like scaling through extended self-similarity analysis. Additionally, we proposed an experimental setup to test our findings. Temporal chaotic statistics were examined using temporal density structure functions, revealing that both weak and strong initial mixing exhibit ESS power-law scaling. 


ESS is traditionally applied to the spatial domain, but our results show that it can also manifest in the time domain, even in systems lacking a clear inertial range in both spatial and time domain. Given the relative simplicity of our setup compared to classical fluid systems, these findings may provide valuable insights into the origins of ESS. 


Spatiotemporal chaos in 1D is closely related to thermalization in integrable or quasi-integrable systems~\cite{kinoshita2006quantum,bland2018probing,thomas2021thermalization,bastianello2020thermalization}, while in higher dimensions, it is associated with turbulence~\cite{zhao2024kolmogorov,madeira2020quantum,navon2016emergence,nazarenko2011wave}. Our findings demonstrate that density structure functions, combined with extended self-similarity, provide an experimentally accessible approach to understanding these complex systems.

\section*{Acknowledgements}
The author thanks Ian Spielman for helpful discussion. 

This work was partially supported by the National Institute of Standards and Technology; the Quantum Leap Challenge Institute for Robust Quantum Simulation (OMA-2120757).

\begin{appendix}
\numberwithin{equation}{section}

\section{Time-Splitting Crank-Nicholson Finite Difference Method}
\label{app:CNFD}
We first split the Hamiltonian into two parts:
\begin{equation}
    H_1 = \frac{1}{2}x^2 + g |\psi|^2, \quad H_2 = -\frac{1}{2} \frac{\partial^2}{\partial x^2}.
\end{equation}
The GPE is then split into two equations that are evolved interchangeably:
\begin{equation}
    i \frac{\partial \psi}{\partial t} = H_1 \psi, \quad i \frac{\partial \psi}{\partial t} = H_2 \psi.
\end{equation}

At each discrete time step $t_n = n \Delta t$, we first evolve the wavefunction using $H_1$ explicitly:
\begin{equation}
    \psi^{n + 1/2} = e^{-i \Delta t H_1} \psi^n.
\end{equation}

Then, we evolve with $H_2$ using the semi-implicit Crank-Nicholson scheme:
\begin{equation}
    \frac{\psi^{n + 1} - \psi^{n + 1/2}}{-i \Delta t} = \frac{1}{2} H_2 \left( \psi^{n + 1} + \psi^{n + 1/2} \right).
\end{equation}
This can be rearranged as:
\begin{equation}
    \psi^{n + 1} = \frac{1 - i \frac{\Delta t}{2} H_2}{1 + i \frac{\Delta t}{2} H_2} \psi^{n + 1/2}.
\end{equation}

Using a finite difference scheme with grid points $x_j = x_0 + jh$, the second derivative becomes:
\begin{widetext}
\begin{equation}
    \frac{\psi_{j}^{n+1}-\psi_j^{n+1/2}}{-i\Delta_t}=-\frac12\frac{1}{2h^2}\left[(\psi^{n+1}_{j+1}-2\psi_{j}^{n+1}+\psi^{n+1}_{j-1})+(\psi_{j+1}^{n+1/2}-2\psi_j^{n+1/2}+\psi_{j-1}^{n+1/2})\right].
\end{equation}
Let $a = \frac{1}{-i \Delta t}$ and $b = \frac{1}{4 h^2}$. The resulting tridiagonal system is:
\begin{equation}
    b \psi_{j + 1}^{n + 1} + (a - 2b) \psi_j^{n + 1} + b \psi_{j - 1}^{n + 1} = -b \psi_{j + 1}^{n + 1/2} + (2b + a) \psi_j^{n + 1/2} - b \psi_{j - 1}^{n + 1/2}.
\end{equation}
\end{widetext}
This holds for $j = 1, 2, \ldots, N_x - 1$, with boundary conditions $\psi_0 = \psi_{N_x} = 0$. This tridiagonal system is solved using the Thomas algorithm.

The Crank-Nicholson scheme is unconditional stable and achieves second-order accuracy in both time and space, ensuring minimal error for small $\Delta t$ and $h$.~\cite{dautray2012r} 
In this work, we use a spatial grid of $N_x = 4097$ with $x_1 = -16$ and $x_{4096} = 16$, giving a grid spacing $h = 32 / 4095$. The time step is set to $\Delta t = \pi \times 10^{-4}$.

\section{Proof of Kohn Theorem}
\label{app:Kohn}
The center of mass of the system is given by
\begin{equation}
    X(t) = \int_{-\infty}^{+\infty} dx\ \psi^*(x,t)\ {x}\ \psi(x,t).
\end{equation}

Using the GPE and its conjugate, and applying integration by parts, the first time derivative becomes:
\begin{equation}
    \frac{dX}{dt} = \int_{-\infty}^{+\infty} dx\ \psi^*(x,t) \left( \frac{-i\hbar}{m} \right) \partial_x \psi(x,t).
\end{equation}

Taking the second time derivative, the only non-vanishing contribution is from the potential term:
\begin{equation}
    \frac{d^2X}{dt^2} = \int_{-\infty}^{+\infty} dx \left( \frac{1}{2} x^2 \psi^* \, \partial_x \psi + \partial_x \psi^* \, \frac{1}{2} x^2 \psi \right) = -X.
\end{equation}

Thus, the center of mass follows simple harmonic motion with angular frequency $\omega = 1$.

It is worth noting that the result above requires both $\psi$ and $\partial_x \psi$ to vanish at the boundaries. Our numerical scheme enforces only the Dirichlet boundary condition, and we approximate the zero-derivative condition by extending the spatial domain, ensuring that boundary derivatives are negligible.

\section{Intermittency of $S_p^R$ under ESS}
\label{app:intermittency}
\begin{figure}[b!]
\centering
\includegraphics{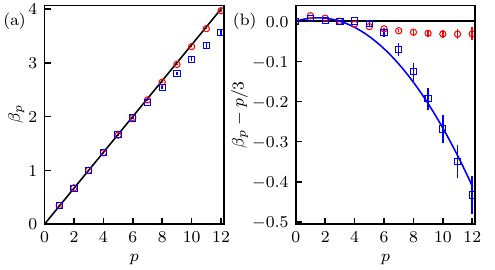}
\caption{Intermittency of $S_p^R$ for the weak mixing ($g=100$, $\alpha=0.1$, red circles) and the strong mixing ($g=100$, $\alpha=1$, blue squares). (a) ESS scaling exponents $\beta_p$, with the black line representing the K41 scaling law, $p/3$. (b) Deviations of $\beta_p$ from $p/3$, with the black horizontal line marking zero deviation. The blue curve shows a parabolic fit based on the KO62 model.}
\label{fig:intermittency}
\end{figure}

To evaluate the intermittency, we calculate the spatial density structure function $S_p^R$ up to $p=12$ for the case shown in Fig.~\ref{fig:ESS}. The scaling exponent $\beta_p$ is plotted in Fig.~\ref{fig:intermittency}(a), and the deviation from K41 scaling $p/3$ is shown in panel (b). For the weak mixing case ($\alpha=0.1$, red circles), the values of $\beta_p - p/3$ appear to saturate, indicating very weak intermittency.

The blue curve represents the KO62 model $\beta_p - p/3 = \kappa p(p-3)$, applied to the strong mixing case ($\alpha=1$, blue squares), with $\kappa = -3.8(1) \times 10^{-3}$ as the intermittency exponent. However, the fit does not entirely capture the observed behavior, suggesting that the system may exhibit multi-fractal characteristics beyond the assumptions of the KO62 model.

\section{Intermittency of $S_p^t$ under ESS}
\label{app:intermittency2}

We compute the temporal density structure function $S_p^t$ up to $p=12$ for the case shown in Fig.~\ref{fig:Sp_thermalize}. 
\begin{figure}[b!]
\centering
\includegraphics{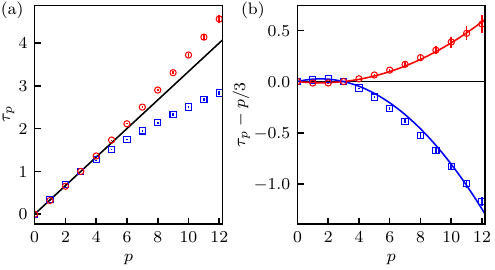}
\caption{Intermittency of $S_p^t$ for weak mixing ($g=100$, $\alpha=0.1$, red circles) and strong mixing ($g=100$, $\alpha=1$, blue squares). (a) ESS scaling exponents $\tau_p$, with the black line representing the K41 scaling law, $p/3$. (b) Deviations of $\tau_p$ from $p/3$, with the black horizontal line indicating zero deviation. Parabolic fits based on the KO62 model are shown by the red and blue curves.}
\label{fig:intermittency_time}
\end{figure}
The scaling exponent $\tau_p$ is displayed in Fig.~\ref{fig:intermittency_time}(a), with the deviation from the K41 scaling $p/3$ shown in panel (b). For the weak mixing case ($\alpha=0.1$, red circles), $\tau_p - p/3$ is positive—an uncommon observation in typical turbulence data. In the strong mixing case ($\alpha=1$, blue squares), the deviation is negative. Both deviations fit well with the KO62 model, $\tau_p - p/3 = \kappa p(p-3)$, as indicated by the curves in panel (b), where $\kappa = 5.40(7) \times 10^{-3}$ for $\alpha=0.1$ and $\kappa = -1.15(2) \times 10^{-2}$ for $\alpha=1$.

\end{appendix}

%

\end{document}